\documentclass[aps,prb,onecolumn,amsmath,amssymb,floatfix,12pt]{revtex4}

\usepackage{graphicx}
\usepackage{dcolumn}
\usepackage{bm}
\usepackage{epstopdf}
\begin{document}

\title{On the Ruderman-Kittel-Kasuya-Yosida Interaction in Graphene}
\author{M. Sherafati}
\author{S. Satpathy}

\affiliation{%
Department of Physics $\&$ Astronomy, University of Missouri,
Columbia, MO 65211, USA}%


\begin{abstract}

{\bf Abstract.} The two dimensionality plus the linear band structure of graphene leads to new behavior of the Ruderman-Kittel-Kasuya-Yosida (RKKY) interaction, which is the interaction between two magnetic moments mediated by the  electrons of the host crystal. We study this interaction  from linear response theory. There are two equivalent methods both of which may be used for the calculation of the susceptibility, one involving the integral over a product of two Green's functions and the second that involves the excitations between occupied and unoccupied states, which was followed in the original work of Ruderman and Kittel. Unlike the $J \propto (2k_FR)^{-2} \sin (2k_FR) $ behavior of an ordinary two-dimensional (2D) metal, $J$ in graphene falls off as $1/R^3$, shows the $1 + \cos ((\bm{K}-\bm{K'}).\bm{R})$-type of behavior, which contains an interference term between the two Dirac cones, and it oscillates for certain directions and not for others. Quite interestingly, irrespective of any oscillations, the RKKY interaction in graphene is always ferromagnetic for moments located on the same sublattice and antiferromagnetic for moments on the opposite sublattices, a result that follows from particle-hole symmetry. \\


\end{abstract}


\maketitle
\pagestyle{empty}

\section{Introduction}

\begin{figure}
\includegraphics[angle=0,width=0.35    \linewidth]{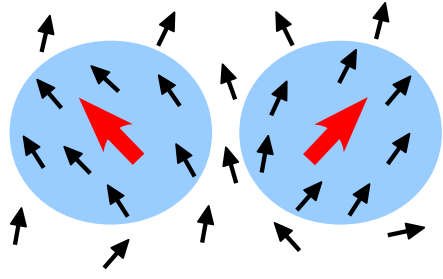}
\caption{\fontsize{10}{10}\selectfont(Color online) The RKKY interaction between two magnetic moments mediated by the host electrons in the crystal.
}
\label{RKKY}
\end{figure}

Graphene, which is a plane of carbon atoms with a honeycomb lattice, is of considerable interest\cite{Castro Neto} owing to its two-dimensionality and a linear band structure as opposed to the quadratic band structure in typical materials. These features also introduce new behaviors in the RKKY interaction, which has been extensively studied beginning with the original works of Ruderman and Kittel\cite{Ruderman}, Kasuya\cite{Kasuya}, and Yosida.\cite{Yosida}
The RKKY interaction is the interaction between two magnetic moments mediated by the conduction electrons in the host material. The first moment perturbs the conduction electrons, which is seen by the second moment leading to an indirect exchange interaction as illustrated in Fig. (\ref{RKKY}).
For the free electron gas with quadratic bands $E = \hbar^2 k^2 / 2 m$ and Fermi momentum $k_F$, the strength of the interaction, $J$, is given by the  expression\cite{Ruderman, Kasuya, Yosida,RKKY-1D,RKKY-2D}
\begin{eqnarray}
J(R)\propto\begin{cases}
\pi / 2 - \text {Si} (x) & \hspace{20mm} \text{(1D) }  \\
\sin x / x^2 & \hspace{20mm} \text{(2D) }  \\
 (x \cos x - \sin x ) / x^4 & \hspace{20mm} \text{(3D) }  \\
\end{cases}
\label{JR}
\end{eqnarray}
where $\text {Si} (x)$ is the sine integral function and $ x = 2 k_F R$ with $R$ being the separation distance between the two moments.
As can be seen from Eq. (\ref{JR}), the common behavior of $J$ in any dimension is characterized by the power-law decay with some oscillations whose period is scaled by the Fermi momentum $k_F$.

In contrast to these simple forms, the form of the RKKY interaction in graphene is quite complex and has been the subject of many papers both for the doped and undoped cases.\cite{MSashi, MSashi2, Vozmediano, Dugaev, Brey, Saremi, Bunder, Black-Schaffer, Kogan} Owing to the lattice structure and a gapless density of states at the Fermi energy with the linear bands occurring at two different points (the Dirac points) in the Brillouin zone (BZ), the RKKY interaction depends on the directionality as well as on the sublattice locations of the two magnetic moments and, in addition, it contains an interference term coming from the two Dirac points $\bm{K}$ and $\bm{K'}$. As we show in this article, the net result is
\begin{eqnarray}
J_{AA}(\bm{R}) &=& -C \times  \frac{1 + \cos [(\bm{K}-\bm{K'}) \cdot \bm{R}]}{(R/a)^3}, \\
\label{JAA}
J_{AB}(\bm{R}) &=& 3C \times  \frac{1 + \cos [(\bm{K}-\bm{K'}) \cdot \bm{R}+\pi-2 \theta_{\bm{R}}]}{(R/a)^3},
\label{JAB}
\end {eqnarray}
where $C $ is a constant, $\theta_{\bm {R}}$ is a position-dependent phase angle, and the subscripts in $J_{\alpha \beta}$ indicate the sublattice location of the moments. The detailed method of how to obtain these results was presented in our earlier study, where we used an expression for the susceptibility in terms of an integral over the product of two  Green's functions. In that method, a cut-off function\cite{Saremi,MSashi} was necessary to evaluate the integrals, although quite satisfactorily the final results did not depend on the exact cut-off function.

It is illuminating to evaluate  the RKKY interaction using an alternative expression for the response, expressed in terms of the excitations of the system, a method familiar in the literature from the original Ruderman-Kittel formulation\cite{Ruderman} and one that has been recently applied to graphene as well.\cite{Kogan} It yields  the same results we had obtained before, without necessitating the use of a cut-off function. In addition to the details of the method, we discuss the salient features of the results for the RKKY interaction in graphene.

\section{Susceptibility}

The RKKY interaction is directly proportional to the susceptibility. The response of the charge density, $n$, to a perturbing potential, $V$, may be written in terms of the integral over the unperturbed Green's function
\begin{equation}
\chi_{\alpha \beta} (\bm{r},\bm{r'}) =
- \frac{2}{\pi} \int^{\varepsilon_F}_{-\infty} d\varepsilon \
{\rm Im} [G^0_{\alpha \beta} (\bm{r}, \bm{r'}, \varepsilon)   G^0_{\beta \alpha} (\bm{r'},\bm{r}, \varepsilon)],
\label{chiGG}
\end{equation}
where $ \chi_{\alpha \beta} (\bm{r},\bm{r'}) \equiv  \delta n_\alpha(\bm{r}) / \delta V_\beta(\bm{r'})$ is the charge susceptibility for a crystal with the Greek subscripts indicating the sublattice indices, $\delta V_\beta(r')$ is a spin-independent perturbing potential and $\delta n_\alpha(\bm{r})$ is the induced charge density that includes both spin channels.

We outline briefly the derivation of the alternative expression for the susceptibility in terms of the energy excitations.
This may be obtained by using the spectral representation of the Green's function
\begin{equation}
G^0_{\alpha \beta} (\bm{r},\bm{r'},\varepsilon)= \sum_{\bm{k}s} \frac{\psi^{\alpha}_{\bm{k}s}(\bm{r})\psi^{*\beta}_{\bm{k}s}(\bm{r'})}{\varepsilon+i\eta - \varepsilon_{\bm{k}s}},
\label{GFspct}
\end{equation}
where $\psi^{\alpha}_{\bm{k}s}$ is the sublattice component of the unperturbed eigenfunction with the corresponding energy $\varepsilon_{\bm{k}s}$. For a crystalline structure, $\{ \bm{k},s \}$ denotes the Bloch momentum and the band index; else, it just denotes a complete set of states. Plugging Eq. (\ref{GFspct}) into the expression Eq. (\ref{chiGG}), one finds the result

\begin{eqnarray}
\chi_{\alpha \beta} (\bm{r},\bm{r'}) =- \frac{2}{\pi} \int^{\varepsilon_F}_{-\infty} d\varepsilon &\times& \nonumber \\
 \sum_{\substack{{\bm{k}s} \\{\bm{k'}s'}}}  \{  {\rm Re} \ [\psi^{\alpha}_{\bm{k}s}(\bm{r})\psi^{* \beta}_{\bm{k}s}(\bm{r'})   \psi^{\beta}_{\bm{k'}s'}(\bm{r'})\psi^{* \alpha}_{\bm{k'}s'}(\bm{r})] \ &{\rm Im}& \ [(\varepsilon+i\eta -\varepsilon_{\bm{k}s})(\varepsilon+i\eta -\varepsilon_{\bm{k'}s'}) ]^{-1}
 \nonumber \\
+ {\rm Im} \ [\psi^{\alpha}_{\bm{k}s}(\bm{r})\psi^{* \beta}_{\bm{k}s}(\bm{r'})\psi^{\beta}_{\bm{k'}s'}(\bm{r'})\psi^{* \alpha}_{\bm{k'}s'}(\bm{r})] \  &{\rm Re}& \ [(\varepsilon+i\eta -\varepsilon_{\bm{k}s})(\varepsilon+i\eta -\varepsilon_{\bm{k'}s'}) ]^{-1} \} .
\end{eqnarray}
It can be easily shown that under the
 interchange of $\bm{k}s$ and $ \bm{k'}s'$, the real part of the product of the four wave functions appearing in the equation above is even, while its imaginary part is odd, and at the same time, both the real and the imaginary parts of the product of the momentum-space Green's function are even. This makes the second line zero. In addition, in order to produce a final compact equation, we replace the real part in the first line by the entire complex quantity, as the extra term introduced thereby gives a zero net result when summed.
 We therefore obtain the expression
\begin{equation}
\chi_{ \alpha \beta} (\bm{r},\bm{r'}) =\sum_{\substack{{\bm{k}s} \\{\bm{k'}s'}}}
\psi^{\alpha}_{\bm{k}s}(\bm{r})\psi^{* \beta}_{\bm{k}s}(\bm{r'})\psi^{\beta}_{\bm{k'}s'}(\bm{r'})\psi^{* \alpha}_{\bm{k'}s'}(\bm{r})  \
\chi(\bm{k}s,\bm{k'}s'),
\label{chiE1}
\end{equation}
where
\begin{equation}
 \chi(\bm{k}s,\bm{k'}s') \equiv -\frac{2}{\pi} \int^{\varepsilon_F}_{-\infty} d\varepsilon \
{\rm Im} [(\varepsilon+i\eta-\varepsilon_{\bm{k}s})(\varepsilon+i\eta-\varepsilon_{\bm{k'}s'})]^{-1}
= 2 \int_{-\infty}^{\varepsilon_F} d\varepsilon \ \big[ \frac{\delta(\varepsilon -\varepsilon_{\bm{k'}s'})}  {\varepsilon -\varepsilon_{\bm{k} s}} +  \frac{\delta(\varepsilon-\varepsilon_{\bm{k}s})}  {\varepsilon-\varepsilon_{\bm{k'}s'}}  \big ] .
\label{chikk'}
\end{equation}
The last equality is obtained by using the relationship $\lim_{\eta \rightarrow 0^+} (x\pm i \eta)^{-1}=\mathcal{P}(1/x) \mp i \pi \delta(x)$.
Clearly, the integral is non-zero only if the eigenstate $\bm {k}s$ is occupied while $\bm {k'}s'$ is empty or vice versa. Thus, corresponding to these two processes,  Eqs. (\ref{chiE1}) and (\ref{chikk'}) lead to two terms,
%
%
which may be combined into a compact expression by using the Fermi  function $f(\varepsilon)=\theta(\varepsilon_F-\varepsilon)$,  where the step function $\theta(x)$ is, as usual, $1$ if $x>0$ and $0$ otherwise. This leads to our desired result
\begin{equation}
\chi_{\alpha \beta} (\bm{r},\bm{r'}) =2 \displaystyle \sum_{\substack{\bm{k},s \\ \bm{k'}s'}} \frac{f(\varepsilon_{\bm{k}s})-f(\varepsilon_{\bm{k'}s'})}{\varepsilon_{\bm{k}s}-\varepsilon_{\bm{k'}s'}}\ \psi^{\alpha}_{\bm{k}s}(\bm{r})\psi^{* \beta}_{\bm{k}s}(\bm{r'})\psi^{\beta}_{\bm{k'}s'}(\bm{r'})\psi^{* \alpha}_{\bm{k'}s'}(\bm{r}).
\label{chiE}
\end{equation}
This is a well-known formula in the linear response theory and is the central equation in this paper, from which we will compute the sublattice susceptibilities for graphene. Note that under the interchange of $\bold{k}s$ and $\bold{k'}s'$, the summand in Eq. (\ref{chiE}) goes into its complex conjugate, so that only the real part survives in the summation.

As already stated, the RKKY interaction can be expressed in terms of the susceptibility.
Taking the interaction between the localized moments and the conduction electrons as a contact interaction in the form
\begin{equation}
V = - \lambda \ ( \bm {S}_1 \cdot \bm {s}_1 + \bm {S}_2 \cdot \bm {s}_2 ),
\label{hamilint}
\end{equation}
where $\bm {s}_i$ is the conduction electron spin density at  site $i$,
the interaction energy between the two localized moments may be written as\cite{Grosso}
\begin{equation}
E(\bm{r},\bm{r'}) = J_{\alpha \beta} (\bm{r},\bm{r'}) \bm {S}_1 \cdot \bm {S}_2,
\label{RKKYE}
\end{equation}
where $ \bm{r},\bm{r'}$  denote the lattice positions of the two spins and the RKKY interaction $J_{\alpha\beta} (\bm{r},\bm{r'}) $ is simply proportional to the susceptibility
\begin{equation}
J_{\alpha \beta} (\bm{r},\bm{r'}) = \frac{\lambda^2 \hbar^2 }{4} \chi_{\alpha \beta} (\bm{r},\bm{r'}).
\label{RKKYJ}
\end{equation}


\begin{figure} [b]
\includegraphics[width=5.0cm]{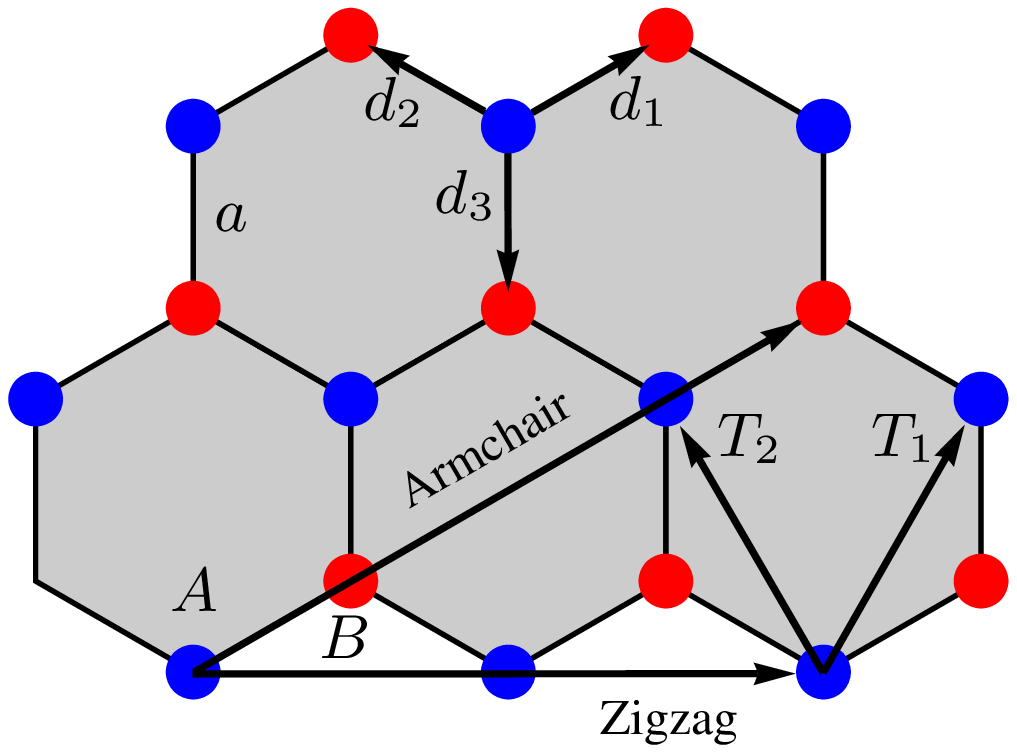}
\hspace{1cm}
\includegraphics[width=2.5cm]{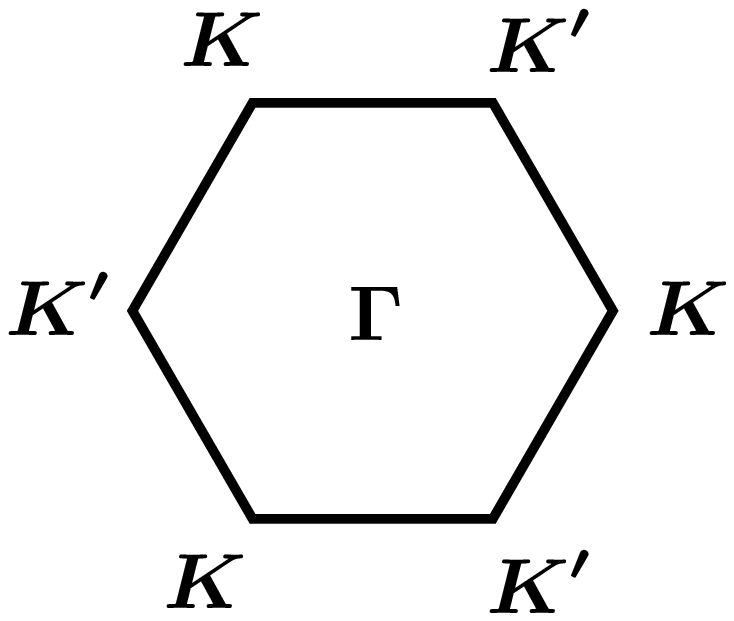}
\caption{\fontsize{10}{10}\selectfont(Color online) The graphene honeycomb lattice with two different sublattices, shown as red and blue dots. The figure also shows the corresponding BZ with the Dirac points $\bm{K}$ and $\bm{K'}$ and two common directions in the direct lattice (zigzag and armchair). $\bm{T}_1$ and $\bm{T}_2$ are the two primitive translation vectors of the direct lattice and the three nearest-neighbor distance vectors are indicated by $\bm{d}_1$, $\bm{d}_2$, and $\bm{d}_3$.
}
\label{graphene}
\end{figure}

\section{The RKKY problem in graphene}

We consider the nearest-neighbor  tight-binding Hamiltonian for the $\pi$ electrons in graphene with the interaction between the host electrons and the  localized magnetic moments given by Eq. (\ref{hamilint}) as before. Thus we have
\begin{equation}
{\cal H}_0=-t \sum_{\langle ij \rangle \sigma} c^{\dagger}_{i\sigma}c_{j\sigma} + H. c.,
\label{hamil0}
\end{equation}
in which $\langle ij \rangle$ denotes summation over distinct pairs of nearest-neighbor sites with hopping parameter $t$ and  $c^{\dagger}_{i\sigma}$ ($c_{i\sigma}$) is the creation (annihilation) operator for an electron with spin index $\sigma$ and combined site-sublattice index $i$. Two set of Bloch sums in the momentum-sublattice representation, viz., $| \bm{k} \alpha\rangle \equiv c^{\dagger}_{\bm{k} \alpha} |0\rangle$ are introduced to diagonalize ${\cal H}_0$.
The Block sums are
\begin{equation}
c^{\dagger}_{\bm{k}\alpha}= N^{-1/2} \sum_i  e^ {i\bm{k}. (\bm{R}_i+\bm{\tau}_{\alpha} )}     c^{\dagger}_{i \alpha},
\label{Bloch}
\end{equation}
 where $N$ is the number of unit cells in the lattice,  $\bm{R}_i$ denotes the cell positions, and $\bm{\tau}_{\alpha}$ denotes the basis atom positions in the unit cell. We take $\bm{\tau}_A = 0$ and $\bm{\tau}_B = \bm{d}_1$.
In the basis of the sublattice Bloch wave functions $| \bm{k} \alpha\rangle $, the Hamiltonian becomes
\begin{equation}
{\cal H}_{\bm{k}}=
\left( \begin{array}{cc}
0 &  f(\bm{k})\\
f^*(\bm{k}) & 0
\end{array}\right),
\label{Hk}
\end{equation}
where  $f(\bm{k}) = -t \ \sum_{j=1}^{3} e^{i \bm{k} \cdot  \bm{d}_j}$, with $\bm{d}_j$ being the nearest-neighbor position vectors and $a$ being the carbon-carbon bond length.
The unperturbed eigenstates of ${\cal H}_{\bm{k}}$ are given by
\begin{eqnarray}
\varepsilon_{\bm{k}s}
                               &=&  s|f(\bm{k})|  \nonumber  \\
\Psi_{\bm{k}s} ^0 &=& \frac{1}{\sqrt 2}
 \left(
 \begin{array}{c}
s e^{i \theta_{\bm{k}}} \\ 1
\end{array}
\right),
\label{estate}
\end{eqnarray}
where the band index $s=\pm$ denotes the conduction and the valence bands and the phase factor $\phi (\bm{k}) \equiv  e^{i \theta_{\bm{k}}}= f(\bm{k}) /   |f(\bm{k})|$. Note that by choosing the Bloch sum (\ref{Bloch}) that includes the phase factor $e^{i \bm{k} \cdot \bm {\tau}_\alpha }$, the basis  wave function for the two atoms in a particular cell also contains the phase factors $e^{i \bm{k} \cdot \bm {\tau}_\alpha } $.

\begin{figure} [b]
\centering
\includegraphics[width=6.0cm]{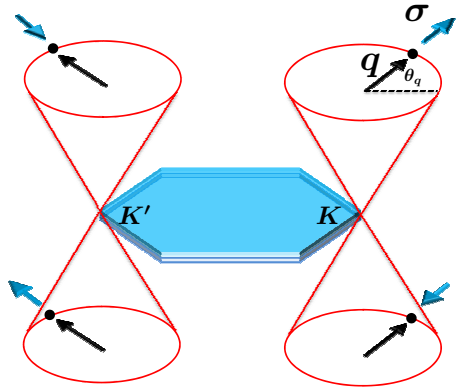}
\caption{\fontsize{10}{10}\selectfont (Color online) The direction of pseudo-spin around Dirac points $\bm{K}$ and $\bm{K'}$. The wave functions have definite helicities $\pm 1$ depending on whether the pseudo-spin $\bm{\sigma}$ is parallel or antiparallel to the  momentum  $\bm{q}$.
}
\label{pseudospin}
\end{figure}
Many of the fascinating properties of graphene like high electron mobility are attributed to its cone-shaped, linear band structure around the corners of the BZ, called the Dirac cones.
The expansion of function $f(\bm{k}=\bm{q}+\bm{K}_D)$ for small  $\bm{q}$ around all six Dirac points takes the form $f(\bm{q}+\bm{K}_D) = v_F q \ \phi (q)$,  leading to the linear band structure
\begin{equation}
\varepsilon = \pm v_F q
\end{equation}
with  Fermi velocity $v_F=3ta/2$. The phase factors appearing in the wave function Eq. (\ref{estate}) are, however, different near different Dirac points.
For all six Dirac points shown in Fig. (\ref{graphene}), starting from the top points and going counter-clockwise, these phase are: $\phi(q) = e^{i(\pi/3-\theta_{\bm{q}})}, \   -e^{i(\pi/3 + \theta_{\bm{q}})}, \  -e^{- i \theta_{\bm{q}}}, \  e^{i(2 \pi/3+\theta_{\bm{q}})}, \  -e^{i(2 \pi/3 - \theta_{\bm{q}})}, \  e^{ i \theta_{\bm{q}}} $ with polar angle of $\bm{q}$ defined as $\theta_{\bm{q}}=\tan^{-1}(q_y/q_x)$.\cite{MSashi}
Note that these phases are a direct consequence of using the extra phase factor $e^ {i\bm{k}.\bm{\tau}_{\alpha}}$ in the Bloch sum, Eq. (\ref{Bloch}). This choice is preferred as all physical quantities will be evaluated at the actual positions of the atoms rather than the  unit cell positions $\bm{R_i}$ in which a particular atom is located. However, the second choice of Bloch sums, not adopted in this paper,  will have the same phase $\phi (q)$ near all Dirac points $\bm{K}$ or $\bm{K'}$ and is used sometimes in the literature.

Near the Dirac cones, the Hamiltonian assumes a simple form
\begin{equation}
{\cal H}_{\bm{q}+\bm{K}}=v_F  \bm{\sigma}^* \cdot \bm{q}, \hspace{10mm}  \
{\cal H}_{\bm{q}+\bm{K'}}=-v_F  \bm{\sigma} \cdot \bm{q},
\label{HKD}
\end{equation}
leading to its interpretation in terms of the pseudo-spins.
Here $\bm{\sigma}=(\sigma_x,\sigma_y)$ are the pseudo-spin Pauli matrices describing the two sublattices and
$\bm{\sigma} ^* = (\sigma_x, -\sigma_y)$. The two-component central-cell wave functions are
\begin{eqnarray}
\Psi^0_{\bm{K'} \pm} = \frac{1}{\sqrt 2}
 \left( \begin{array}{c}
\mp e^{-i \theta_{\bm{q}}} \\
1
\end{array}\right) , \
\Psi^0_{\bm{K} \pm} = \frac{1}{\sqrt 2}
\left( \begin{array}{c}
\pm e^{i \theta_{\bm{q}}} \\
1
\end{array}\right),
\label{PsiE}
\end{eqnarray}
in the basis set of the carbon orbitals without any phase factors $e^{i \bm{k} \cdot \bm {\tau}_\alpha } $ in their definitions.
The wave functions have definite helicities $\pm 1$ corresponding to the eigenvalues of the operator $ \hat{h} =  \bm{\sigma} \cdot  \bm{\hat {q}}$ as indicated in Fig (\ref{pseudospin}).


\subsection{Moments on the Same Sublattice: $J_{AA}(\bm{R})$}

\begin{figure}
\begin{@twocolumnfalse}
\includegraphics[width=7.5cm]{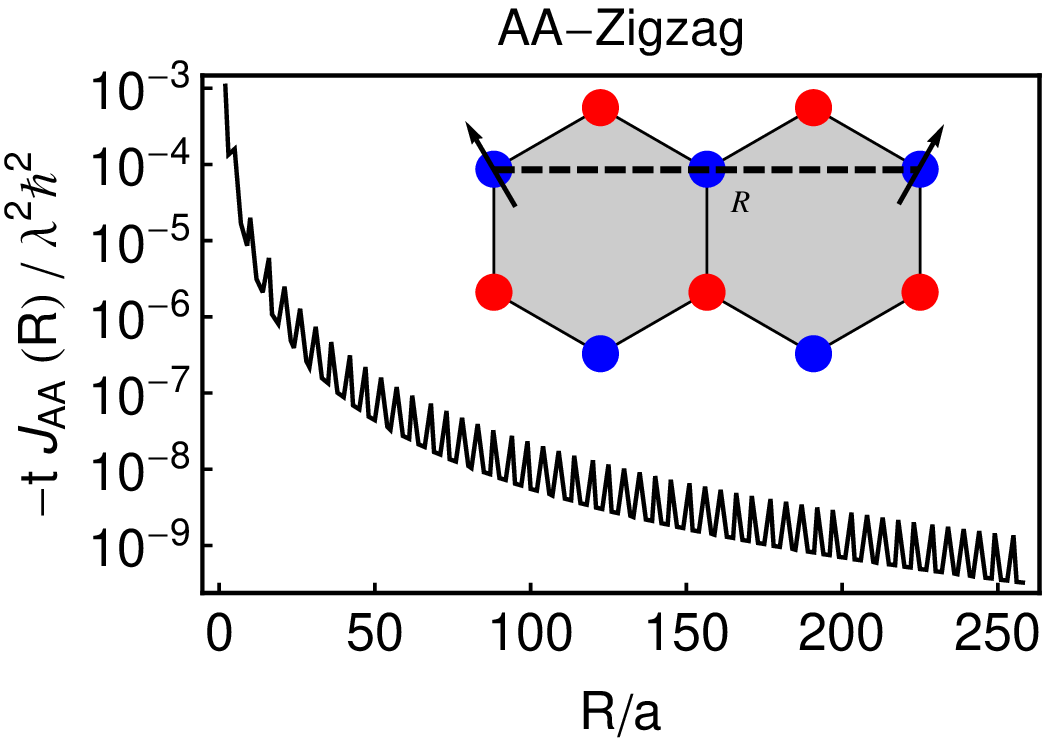}
\hspace{5mm}
\includegraphics[width=7.5cm]{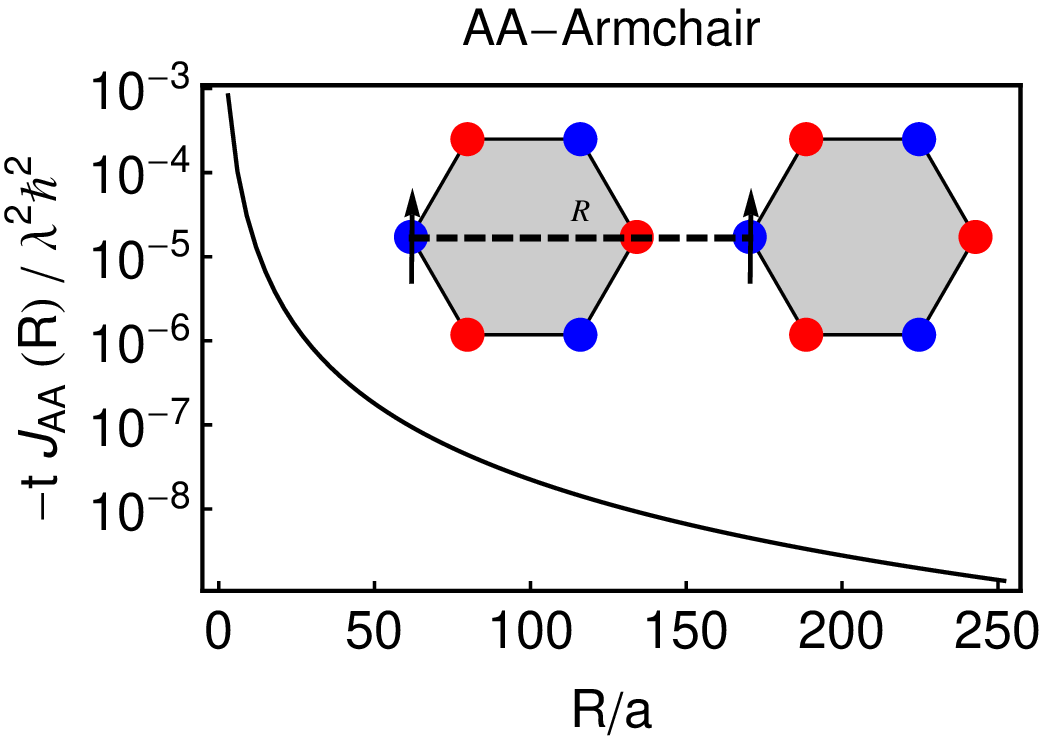}
\caption{\fontsize{10}{10}\selectfont (Color online) RKKY interaction $J_{AA}$ between two moments on the same sublattice located either along the zigzag or the armchair directions as obtained from Eq. (\ref{JAA}). Note that $J_{AA}$ is negative for all $R$ indicating a ferromagnetic interaction. Consistent with Eq. (\ref{JAA}), $J$ is oscillatory for the zigzag direction and smoothly decays along the armchair direction, while always remaining ferromagnetic.}
\label{Fig-JAA}
\end{@twocolumnfalse}
\end{figure}

We will now use these eigenstates to evaluate the RKKY interaction using the susceptibility expression  Eq. (\ref{chiE}).
Using Eq. (\ref{estate}) in Eq. (\ref{chiE}) for both moments on $A$-sublattice, one at the origin and the other at the  atom position $\bm{R}$, we find
\begin{equation}
\chi_{ A A} (0,\bm{R}) = \frac{1}{N^2} \sum_{\substack{{\bm{k},\bm{k'}} \\{\varepsilon_{\bm{k}-}< \varepsilon_F  <\varepsilon_{\bm{k'}+}}}}
 \frac{e^{-i (\bm{k}-\bm{k'}) \cdot \bm{R}}}{\varepsilon_{\bm{k}-}-\varepsilon_{\bm{k'}+}}.
\label{chiAA1}
\end{equation}
We now evaluate Eq. (\ref{chiAA1}) for the linear Dirac bands. We can construct the BZ such that it encloses two top Dirac points shown in Fig (\ref{graphene}) and perform the summations over $\bm{k}$ and $\bm{k'}$ over two circles centered at the Dirac points. We first perform the $\bm{k'}$ summation, which yields
\begin{equation}
\frac{1}{N} \sum_{\bm{k'}} \frac{e^{i \bm{k'} \cdot \bm{R}}}{\varepsilon_{\bm{k}-}-\varepsilon_{\bm{k'}+}}=
e^{i \bm{K} \cdot \bm{R}} \big( \frac{1}{N} \sum_{\bm{q}_1} \frac{e^{i \bm{q}_1 \cdot \bm{R}}}{\varepsilon_{\bm{k}-}-v_Fq_1}\big)+e^{i \bm{K'} \cdot \bm{R}} \big(\frac{1}{N} \sum_{\bm{q}_2} \frac{e^{i \bm{q}_2 \cdot \bm{R}}}{\varepsilon_{\bm{k}-}-v_Fq_2}\big),
\label{chiAA2}
\end{equation}
where $\bm {q}_1, \bm{q}_2 $ denote the momentum with respect to the two Dirac points.
Using the Jacobi-Anger expansion\cite{GRHandbook} for the exponentials, viz.,
\begin{equation}
e^ {\pm i \bm{q} \cdot \bm{R}} = J_0 (qR) + 2 \sum_{n=1}^\infty (\pm i)^n J_n (qR) \cos [n (\theta_{\bm{q}} -\theta_{\bm{R}})],
\label{Jacobi}
\end {equation}
where $J_n(x)$ is the Bessel function and the integral $\int^{2 \pi}_0 e^ {\pm i \bm{q} \cdot \bm{R}} d \theta_{\bm{q}}=2 \pi  J_0 (qR)$, Eq. (\ref{chiAA2}) yields
\begin{equation}
\frac{1}{N} \sum_{\bm{k'}} \frac{e^{i \bm{k'} \cdot \bm{R}}}{\varepsilon_{\bm{k}-}-\varepsilon_{\bm{k'}+}}=
(e^{i \bm{K} \cdot \bm{R}}+e^{i \bm{K'} \cdot \bm{R}}) (\frac{2 \pi}{\Omega_{BZ}}) \int^{q'_c}_0\frac{q' J_0(q'R)}{\varepsilon_{\bm{k}-}-v_Fq'} \ dq'.
\label{chiAA3}
\end{equation}
Here we have used $N^{-1} \displaystyle \sum_{\bm{q}}\rightarrow (2\Omega_{BZ})^{-1}\int d^2q$ with $\Omega_{BZ}$ being the area of the BZ. We now perform the $\bm{k}$-summation in Eq. (\ref{chiAA1}) using Eq. (\ref{chiAA3}) following the similar steps as above and finally get the expression
\begin{equation}
\chi_{ A A} (0,\bm{R}) = -\frac{2}{v_F} (\frac{2 \pi}{\Omega_{BZ}})^2 \frac{1 + \cos [(\bm{K}-\bm{K'}) \cdot \bm{R}]}{R^3} \ I_{AA},
\label{chiAAI}
\end{equation}
\begin{equation}
I_{AA}= \int^\infty_0\int^\infty_0\frac{xx'J_0(x)J_0(x')}{x+x'}dx dx',
\label{IAA}
\end{equation}
where $x=qR,  x'=q'R$, and no cutoff has been used for the Dirac cones.
To evaluate $I_{AA}$ we define the function
\begin{equation}
H(s)= \int^\infty_0\int^\infty_0 e^{-s(x+x')} \frac{xx'J_0(x)J_0(x')}{x+x'}dx dx'.
\label{H1}
\end{equation}
The derivative of $H(s)$  gives the square of a Laplace transform, which can be easily evaluated, with the result
\begin{equation}
\frac{dH(s)}{ds} =  -  \big[   \int^\infty_0 e^{-sx} x J_0(x)\big] ^2
=  -(\mathcal{L}[xJ_0(x)])^2
= -\frac{s^2}{(1+s^2)^3}.
\label{DH}
\end{equation}
Integrating this and determining the constant of integration from  the condition $ H (0) = I_{AA}$ (see Eqs. (\ref{H1}) and
(\ref{IAA})), we find
\begin{equation}
H(s)= \frac{s(1-s^2)}{8 (1+s^2)^2}-\frac{1}{8} \tan^{-1} s+I_{AA}.
\label{H2}
\end{equation}
From the definition of $H(s)$ in Eq. (\ref{H1}) we see that $\lim _{s \rightarrow \infty}H(s)=0$ and evaluating the right hand side for $s = \infty$, we immediately find $I_{AA}=\pi/16$.
Plugging  this result for $I_{AA}$ into Eqs. (\ref{chiAAI}) and (\ref{RKKYJ}), we find the RKKY interaction to be
\begin{equation}
J_{AA}(\bm{R}) = -C \times  \frac{1 + \cos [(\bm{K}-\bm{K'}) \cdot \bm{R}]}{(R/a)^3},
\label{JAA}
\end{equation}
where $C \equiv 9 \lambda^2 \hbar^2/ (256 \pi t)$. As $C>0$, Eq. (\ref{JAA}) represents a ferromagnetic coupling between the moments on the same sublattice. One can simply show that Dirac-cone oscillatory factor, $1 + \cos [(\bm{K}-\bm{K'}) \cdot \bm{R}]$ takes the sequence of triplets of 2, 1/2, 1/2, ... with distance $R$ along the zigzag direction, and becomes always $2$ for the armchair direction. These results are shown in Fig. (\ref{Fig-JAA}). This is consistent with the conclusion that\cite{Saremi}  in the presence of the  particle-hole symmetry (which is true for a bipartite lattice with no interaction between the members of the same sublattices), the RKKY interaction between two moments placed on the same sublattice is ferromagnetic, while those placed on the opposite sublattices is antiferromagnetic. Presence of the second nearest-neighbor interaction breaks this symmetry,
which is relatively weak in graphene.\cite{Nanda}
%
\subsection{Moments on the Opposite Sublattices: $J_{AB}(\bm{R})$}

\begin{figure} [t]
\begin{@twocolumnfalse}
\includegraphics[width=7.5cm]{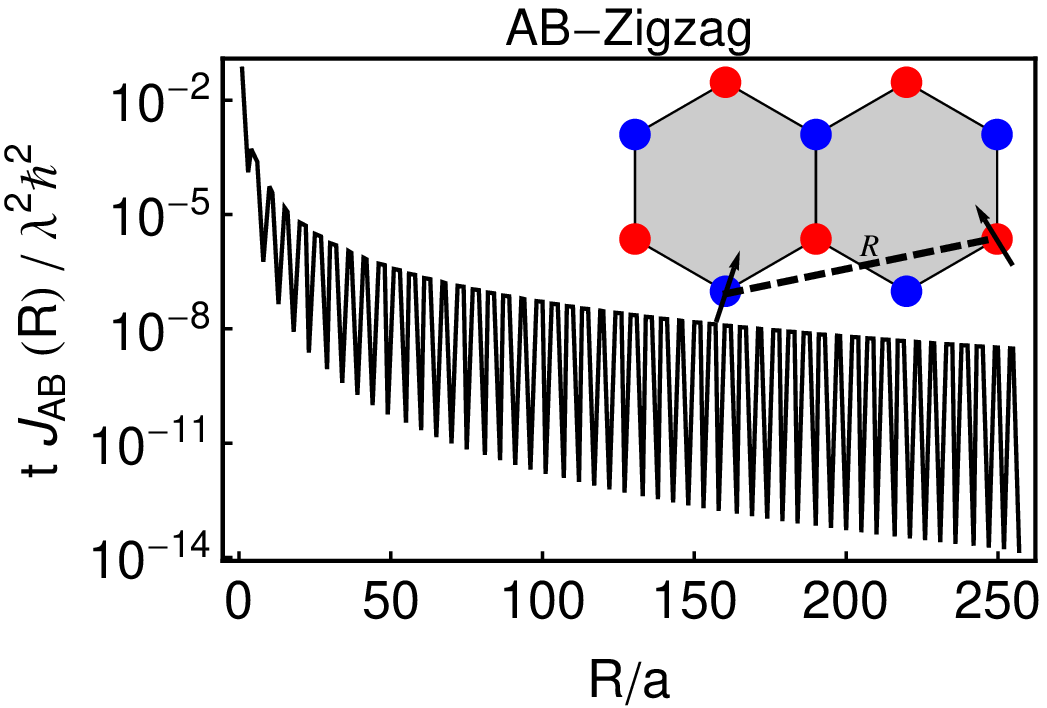}
\hspace{5mm}
\includegraphics[width=7.5cm]{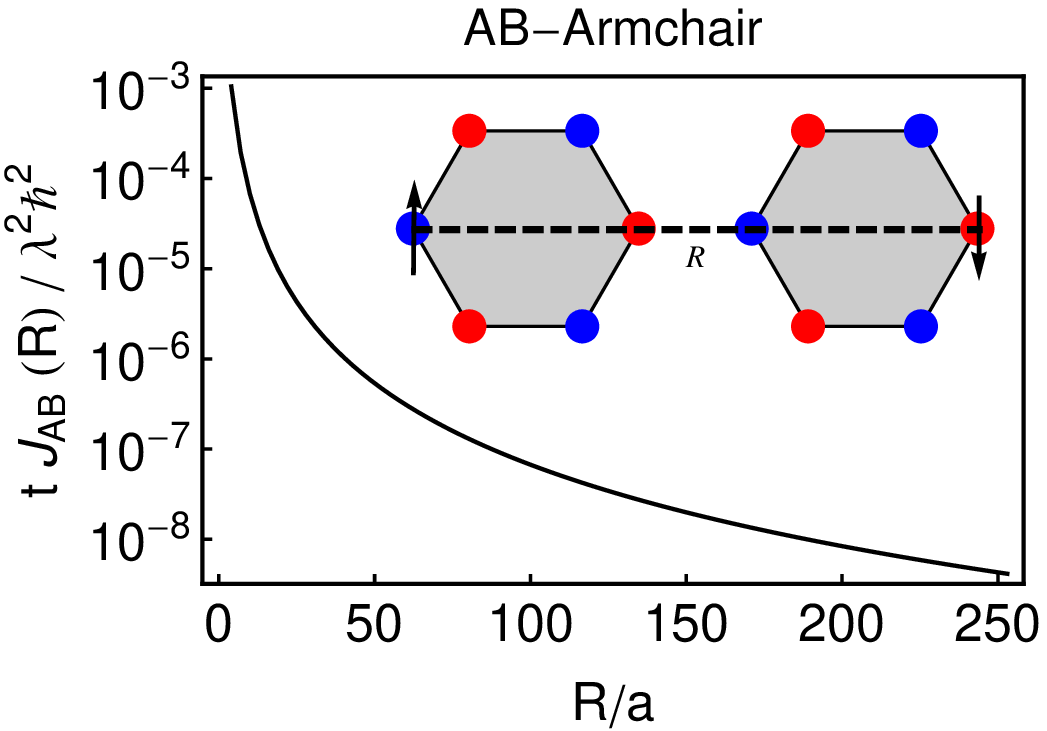}
\caption{\fontsize{10}{10}\selectfont (Color online) $J_{AB}$ for the zigzag direction calculated from Eq. (\ref{JAB}). Note that $J_{AB}$ is positive for all $R$ indicating an  antiferromagnetic coupling  for all distances, even though the magnitude can oscillate.}
\label{Fig-JAB}
\end{@twocolumnfalse}
\end{figure}

For two moments located on the opposite sublattices, the first on the $A$ sublattice atom  at the origin and the second at the atom position $\bm{R}$ on the $B$-sublattice, Eqs. (\ref{chiE}) and (\ref{estate}) yield
\begin{equation}
\chi_{ A B} (0,\bm{R}) =\frac{1}{N^2} \sum_{\substack{{\bm{k},\bm{k'}} \\{\varepsilon_{\bm{k}s}< \varepsilon_F   <\varepsilon_{\bm{k'}s'}}}}   \frac{e^{i (\theta_{\bm{k}}-\theta_{\bm{k'}})} e^{-i (\bm{k}-\bm{k'}) \cdot \bm{R}}}{\varepsilon_{\bm{k}-}-\varepsilon_{\bm{k'}+}} .
\label{chiAB1}
\end{equation}
Following similar algebra as in the previous section and performing the angle integration after the  Jacobi-Anger expansion Eq. (\ref{Jacobi}), viz., $\int^{2 \pi}_0 e^ {\pm i \bm{q} \cdot \bm{R}} e^{\pm i\theta_{\bm{q}}}d \theta_{\bm{q}}=\pm 2 \pi i J_1 (qR)e^{\pm i\theta_{\bm{R}}}$, the susceptibility is given by
\begin{equation}
\chi_{ A B} (0,\bm{R}) = \frac{2}{v_F} (\frac{2 \pi}{\Omega_{BZ}})^2 \frac{1 + \cos [(\bm{K}-\bm{K'}) \cdot \bm{R}+\pi-2 \theta_{\bm{R}}]}{R^3} \ I_{AB},
\label{chiABI}
\end{equation}
\begin{equation}
I_{AB}= \int^\infty_0\int^\infty_0\frac{xx'J_1(x)J_1(x')}{x+x'}dx dx',
\label{IAB}
\end{equation}
where $x=qR$ and $x'=q'R$. This integral can be evaluated following the above method of Laplace transform with replacing
 $J_0(x)$ by $J_1(x)$ in Eqs. (\ref{H1}) and (\ref{DH}). The result is $I_{AB}=3 \pi/16$. With this, Eqs. (\ref{chiABI}) and  (\ref{RKKYJ}) yield the RKKY interaction
\begin{equation}
J_{AB}(\bm{R}) = 3C \times  \frac{1 + \cos [(\bm{K}-\bm{K'}) \cdot \bm{R}+\pi-2 \theta_{\bm{R}}]}{(R/a)^3}.
\label{JAB}
\end{equation}
Clearly, $J_{AB}$ is always antiferromagnetic as required by particle-hole symmetry, even though its magnitude may oscillate with distance. The results are plotted in Fig. \ref{Fig-JAB} for two different directions in the graphene lattice.

\section{Summary}

We have studied the RKKY interaction in graphene by computing the susceptibility Eq. (\ref{chiE}) in terms of the one-particle excitations in the system following the original method of Ruderman and Kittel.\cite{Ruderman}
The results are the same as evaluated using the integration over the product of the Green's function Eq. (\ref{chiGG}), a method we have adopted in our earlier works.\cite{MSashi, MSashi2} The present method is somewhat better for graphene in that no cut-off functions are needed to perform the integrals.

The RKKY interaction in graphene has several interesting features.
Unlike the  $1/R^2$ fall-off of the RKKY interaction in the standard 2D metals with quadratic dispersion, the linear band structure of graphene leads to the $1/R^3$ dependence on distance. The interference between the two Dirac cones in the Brillouin zone produces an interference effect that leads to an oscillatory behavior of the RKKY interaction. However, even though the magnitude may oscillate with distance along certain directions, the sign of the interaction is always ferromagnetic for moments located on the same sublattice and antiferromagnetic for moments located on the opposite sublattices, a result that follows from the particle-hole symmetry of graphene.\cite{Saremi}
 Lately it has been possible to electron or hole dope graphene by a gate voltage. The same analysis can be extended to this case. Some results for the RKKY interaction for the doped system have been presented in the literature.\cite{MSashi2}

 We finally note that there is currently considerable interest in the topological insulators, where one also has two-dimensionality and a linear band structure. However, there are important differences between the topological insulators and graphene. For instance, graphene contains an even number  of Dirac cones (four including spin and valley degeneracies), while the topological insulators contain an odd number of cones. Secondly, in graphene, the linear band structure originates by the presence of a pseudo-spin, while in the topological insulators we have real spins, so that a magnetic impurity opens up a local gap and suppresses the local density of states. As in the usual Fermi liquid, if the surface state has a finite Fermi wave vector $k_F$, the sign of the RKKY interaction oscillates with wavelength $\propto k_F/2$. However, if the Fermi level is close to the Dirac point, the RKKY interaction will always be ferromagnetic as a uniform spin polarization can maximize the gap opened on the surface.\cite{Qi}

\section{Acknowledgement}
This work was supported by the U. S. Department of Energy through Grant No.
DOE-FG02-00ER45818. We thank Jet Foncannon for helpful discussions.

\section{References}

\end{document}